\documentclass[12pt]{article}
\usepackage{amsmath}
\usepackage{amssymb}
\usepackage{graphicx}
\usepackage{bm}
\usepackage{upgreek}

\begin{document}

\begin{center}
{\bf \Large Little Rip cosmological models with time-dependent equation of state}

\bigskip

I. Brevik\footnote{iver.h.brevik@ntnu.no}

\bigskip
Department of Energy and Process Engineering, Norwegian University of Science and Technology, N-7491 Trondhein, Norway

\bigskip
V. V. Obukhov, K. E. Osetrin and A. V. Timoshkin

\bigskip

Tomsk State Pedagogical University, 634061 Tomsk, Russia

\end{center}

\begin{abstract}

Specific dark energy models, leading to the Little Rip (LR) cosmology in the far future, are investigated. Conditions for the occurrence of  LR in terms of the parameters present in the proposed  equation of state for the dark energy cosmic fluid are studied. Estimates about the time needed before the occurrence of the rip singularity in the standard LR model and the model in which the universe approaches the de Sitter space-time  asymptotically, are given.

\end{abstract}

\bigskip

\section{Introduction}
The discovery of the accelerating universe has led to the appearance of new theoretical models (for recent reviews, see \cite{bamba12,nojiri11}). The cosmic acceleration can be explained via the introduction of dark energy (recent references are \cite{li11,cai10}), with strange properties like negative pressure and/or negative entropy. According to the latest supernovae observations the dark energy accounts for 73\% of the total mass energy of the universe (see, for instance, \cite{kowalski08}).

One characteristic feature of the new cosmology is the Big Rip (BR) phenomenon, i.e. a singularity of the universe to be encountered in a finite time. Pioneering papers on the Big Rip were given in \cite{caldwell02,caldwell03,mcinnes02,faraoni02,nojiri03}. The Big Rip  means that one or more of the physical quantities go to infinity in a finite time $t$ in the future. Mathematically, it implies divergent integrals following from the Friedmann equations.

A softer variant of the future singularity phenomenon is the so-called Little Rip (LR), characterized by an energy density $\rho$ increasing with time but in an asymptotic sense, so that an infinite time is required to reach the singularity. The LR scenario was proposed in \cite{frampton11}. It corresponds to an equation of state parameter $w<-1$, but $w\rightarrow -1$ asymptotically. A viscous generalization of the LR theory was recently given in \cite{brevik11}. The energy density grows with time but not fast enough for the occurrence of the Big Rip singularity. The LR models describe intermediate evolutions between an asymptotic de Sitter expansion and a BR evolution.

The so-called Pseudo-Rip (PR), proposed in \cite{frampton12}, is a third variant of the theory. This interesting possible scenario is related to the LR cosmology when the  Hubble parameter tends to infinity in the remote future \cite{frampton12B,astashenok12,astashenok12B,astashenok12C,nojiri11B},
\begin{equation}
H(t) \rightarrow H_\infty <\infty, \quad t\rightarrow \infty. \label{1}
\end{equation}
A model discussing LR in multivariate Gauss-Bonnet theory with dilaton was constructed in \cite{makarenko12}. The LR cosmology may be described in terms of a general fluid with a complicated equation of state \cite{caldwell03,sami04}. The nonsingular LR cosmology may lead to structure disintegration in the future (galaxies, the Sun system, etc), similarly as the BR.

In the present paper we study the influence of the time dependent parameters $w$ and $\Lambda$ in the equation of state (see equation (\ref{5}) below) upon the occurrence of LR/PR in various cosmological models. Estimates for the time required for disintegration of gravitational bound systems are given.

\section{Equation of state with time-dependent $w$ and $\Lambda$ in the Little Rip model}

Let us consider the following LR model with a given Hubble parameter $H$:
\begin{equation}
H(t)=H_0e^{\lambda t}, \quad H_0>0, \quad \lambda >0. \label{2}
\end{equation}
Here it is natural to associate $t=0$ with the present time, so that $H_0$ becomes the present-time Hubble parameter. The Friedmann  equation for a spatially flat universe is
\begin{equation}
\rho=\frac{3}{\kappa^2}H^2, \label{3}
\end{equation}
where $\rho$ is the energy density, $H=\dot{a}/a$ the Hubble parameter, $a(t)$ the scale factor, and $\kappa^2=8\pi G$ with $G$ being Newton's constant. Thus
\begin{equation}
\dot{\rho}=\frac{6\lambda}{\kappa^2}H^2. \label{4}
\end{equation}
We assume now that our universe is filled with an ideal fluid (dark energy) obeying an inhomogeneous equation of state \cite{brevik04}
\begin{equation}
p=w(t)\rho+\Lambda(t), \label{5}
\end{equation}
where $p$ is the pressure and $w(t), \Lambda(t)$ are time-dependent parameters. This form is related to the one used in our earlier paper \cite{brevik07}; cf. also \cite{brevik12}.

The energy conservation law is
\begin{equation}
\dot{\rho}+3H(\rho+p)=0. \label{6}
\end{equation}
Taking into account (\ref{4})-(\ref{6}) we obtain
\begin{equation}
\frac{6\lambda}{\kappa^2}H^2+3H\left[\frac{3}{\kappa^2}[1+w(t)]H^2+\Lambda(t)\right]=0. \label{7}
\end{equation}
Solving equation (\ref{7}) with respect to $w(t)$ we have
\begin{equation}
w(t)=-\frac{2}{3}\frac{\lambda}{H}-\frac{\Lambda(t)\kappa^2}{3H^2}-1. \label{8}
\end{equation}

Let us investigate various cases:

\bigskip
\noindent {\bf 1.} Assume that the parameter $\Lambda$ does not dependent on time, $\Lambda(t)=\Lambda_0$. Then $w(t) \rightarrow -1$ asymptotically  from below, in the far future. Thus we see that for an ideal fluid obeying equations (\ref{5}) and (\ref{8}) the LR scenario is found.

If instead the parameter $w$ does not depend on time, $w(t)=w_0$, we can solve equation (\ref{7}) with respect to $\Lambda(t)$ to get
\begin{equation}
\Lambda(t)=-\frac{H}{\kappa^2}\left[ 2\lambda +3H(1+w_0)\right]. \label{9}
\end{equation}
As $w_0<-1$ in order to correspond to a dark fluid, we see that $\Lambda(t)\rightarrow \infty$ in the far future. The LR behavior in this case is caused by the parameter $w_0$.

\bigskip
\noindent {\bf 2.} Let us now consider instead of (\ref{2}) another LR model \cite{brevik11} where the Hubble parameter is given as
\begin{equation}
H(t)=H_0\,e^{Ce^{\lambda t}}. \label{10}
\end{equation}
Here $H_0, C$ and $\lambda$ are positive constants.
(Note that with this definition, the present-time value of the Hubble constant becomes $H(0)=H_0\,e^C$.) In the paper \cite{frampton11} it was shown that this model may be a realistic one, as it is compatible with the observations.

We now find
\[ \rho=\frac{3}{\kappa^2}H_0^2\,e^{2Ce^{\lambda t}}, \]
\begin{equation}
\dot{\rho}=\frac{6\lambda}{\kappa^2}H^2\ln \frac{H}{H_0}, \label{11}
\end{equation}
and in view of (\ref{10}) and (\ref{11}) equation (\ref{6}) becomes
\begin{equation}
\frac{6\lambda}{\kappa^2}H^2\ln \frac{H}{H_0}+3H\left[ \frac{3}{\kappa^2}[1+w(t)]H^2+\Lambda(t)\right]=0. \label{12}
\end{equation}
Let us suppose that the parameter $\Lambda(t)$ is proportional to the square of the Hubble parameter \cite{houndjo11}, that is
\begin{equation}
\Lambda(t)=\gamma H^2, \label{13}
\end{equation}
with $\gamma$ a constant. Taking (\ref{13}) into account, we solve equation (\ref{12}) with respect to $w(t)$ to obtain
\begin{equation}
w(t)=-\frac{2}{3}\frac{\lambda}{H}\ln \frac{H}{H_0}-\frac{\gamma}{3}\kappa^2-1. \label{14}
\end{equation}
Now writing the parameter $w(t)$ in the form
\begin{equation}
w(t)=-1-\frac{\delta}{H^2}, \label{15}
\end{equation}
with $\delta$ a positive constant, we obtain from (\ref{12})
\begin{equation}
\Lambda(t)=-\frac{2\lambda}{\kappa^2}H\ln \frac{H}{H_0}+\frac{3\delta}{\kappa^2}. \label{16}
\end{equation}
As shown in \cite{frampton11} we can easily find models in which there are more complicated behaviors of $H$, such as
\begin{equation}
H=H_0\exp \left( C_0\exp C_1\left(\exp C_2\left(\exp...\left(C_n\exp(\lambda t)\right)\right)\right)\right), \label{17}
\end{equation}
where $C_0,C_1,...,C_n$ are positive constants. In this case the energy conservation law takes the form
\begin{equation}
\frac{6\lambda}{\kappa^2}H^2\ln \frac{H}{H_0}{\rm Ln} \frac{H}{H_0}+3H\left[ \frac{3}{\kappa^2}[1+w(t)]H^2+\Lambda(t)\right]=0, \label{18}
\end{equation}
where we have defined Ln as
\[ {\rm Ln}\frac{H}{H_0} \equiv \left[ \ln \left(\frac{1}{C_0}\ln \frac{H}{H_0}\right)...\ln\left(\frac{1}{C_{k-1}}...\ln\left(\frac{1}{C_0}\ln \frac{H}{H_0}\right)\right)\right], \]
with $k \in (1,n)$ and $n\in N$.

Using (\ref{13}) to solve equation (\ref{18}) with respect to $w(t)$, we obtain
\begin{equation}
w(t)=-\frac{2}{3}\frac{\lambda}{H}\ln \frac{H}{H_0}{\rm Ln}\frac{H}{H_0}-\frac{\gamma}{3}\kappa^2-1. \label{19}
\end{equation}
Analogously, by generalizing (\ref{16}) we obtain
\begin{equation}
\Lambda(t)=-\frac{2\lambda}{\kappa^2}H\ln \frac{H}{H_0}{\rm Ln} \frac{H}{H_0}+\frac{3\delta}{\kappa^2}. \label{20}
\end{equation}

\bigskip

\noindent {\bf 3.} Now go on to consider a Pseudo-Rip model \cite{frampton12} with a different behavior of $H$ \cite{frampton12B}:
\begin{equation}
H(t)=H_0 -H_1e^{-\lambda t}, \label{21}
\end{equation}
where $H_0, H_1$ and $\lambda$ are positive constants. We assume that $H_0 >H_1$ when $t>0$.  Since the second term decreases with increasing $t$, the universe approaches asymptotically the de Sitter space-time with Hubble constant $H_0$ ($H_0$ here means the same as $H_\infty$ in (\ref{1})).

We now find
\[ \rho=\frac{3}{\kappa^2}(H_0-H_1e^{-\lambda t})^2, \]
\begin{equation}
\dot{\rho}=\frac{6\lambda}{\kappa^2}H(H-H_0). \label{22}
\end{equation}
Using (\ref{18}) we can rewrite the energy conservation equation as
\begin{equation}
\frac{6\lambda}{\kappa^2}H(H-H_0)+3H\left[\frac{3}{\kappa^2}[1+w(t)]H^2+\Lambda(t)\right] =0. \label{23}
\end{equation}
We will now investigate this kind of PR, in analogy with the earlier models (\ref{2}) and (\ref{10}).

First, if we assume $\Lambda(t)=\Lambda_0$, then we find
\begin{equation}
w(t)=-\frac{2}{3}\lambda \frac{H-H_0}{H^2}-\frac{\kappa^2}{3H^2}\Lambda_0-1, \label{24}
\end{equation}
which shows that the PR behavior is determined by the parameter $\Lambda_0$.

Next, if we take $\Lambda(t)=\gamma H^2$, we find
\begin{equation}
w(t)=-\frac{2}{3}\lambda \frac{H-H_0}{H^2}-\frac{\kappa^2\gamma}{3}-1, \label{25}
\end{equation}
showing that the PR is connected with the Hubble parameter (\ref{21}).

Solving $\Lambda(t)$ from (\ref{23}) and taking $w(t)=w_0$, we see that the PR is determined by the parameter $w_0$:
\begin{equation}
\Lambda(t)=-\frac{2\lambda}{\kappa^2}(H-H_0)-\frac{3}{\kappa^2}(1+w_0)H^2. \label{26}
\end{equation}
In view of (\ref{15}) this means
\begin{equation}
\Lambda(t)=-\frac{2\lambda}{\kappa^2}(H-H_0)-\frac{3}{\kappa^2}\delta. \label{27}
\end{equation}
Thus, we have presented the appearance of LR and PR from the equation of state (\ref{5}).

\section{The inertial force in Little Rip cosmology}

During the universe expansion the relative acceleration between two points whose distance is $l$ is equal to $l \ddot{a}/a$, where $a$ is the scale factor. A particle with mass $m$ at a given point will be subject to an inertial force \cite{frampton12}
\begin{equation}
F_{\rm in}=ml\frac{\ddot{a}}{a}=ml(\dot{H}+H^2). \label{28}
\end{equation}
Let us assume that two particles are bound by a constant gravitational force $F_0$. If $F_{\rm in}>0$ and $F_{\rm in}>F_0$, the particles become unbound. This is the Big Rip phenomenon caused by the accelerating expansion; the Sun system or the galaxies in the universe may become gravitationally unbound. It is convenient to define the dimensionless parameter \cite{frampton12,astashenok12}
\begin{equation}
\bar{F}_{\rm in}=\frac{2\rho(a)+\rho'(a)a}{\rho_0} = 6\frac{\ddot{a}}{a\rho_0},  \label{29}
\end{equation}
where $\rho_0$ is the dark energy density at present.

Now consider the LR model as determined by (\ref{2}). The inertial force is \cite{frampton12B}
\begin{equation}
F_{\rm in}=ml(\lambda H_0e^{\lambda t}+H_0^2e^{2\lambda t}). \label{30}
\end{equation}
Hence, at time $t\rightarrow +\infty$ the inertial force $F_{\rm in}\rightarrow +\infty$. This characterizes the LR; under certain conditions a disintegration may occur for gravitational force structures.

Let us investigate the influence from the parameters $w$ and $\Lambda$ in the equation of state for dark energy upon the time $t_{LR}$ needed for disintegration in the LR model governed by (\ref{2}). In the equation of state (\ref{5}) we take
\[ w(t)=w_0, \]
\begin{equation}
w_0<-1, \label{31}
\end{equation}
\[ \Lambda(t)=\Lambda_0, \]
where $w_0$ and $\Lambda_0$ are constants.

Assume now that
\begin{equation}
w_0=-1-\frac{\lambda^2}{3\Lambda_0\kappa^2}. \label{32}
\end{equation}
The energy conservation law becomes
\begin{equation}
(\lambda H_0)^2e^{2\lambda t}-2\Lambda_0\kappa^2\lambda H_0e^{\lambda t}-\Lambda_0^2\kappa^4=0. \label{33}
\end{equation}
From this we find the disintegration time $t_{LR}$:
\begin{equation}
t_{LR}=\frac{1}{\lambda}\ln \frac{(1+\sqrt{2})\Lambda_0\kappa^2}{\lambda H_0}, \label{34}
\end{equation}
Parameter estimates show that the Rip takes place after some billions of years, and that the Sun-Earth system disintegrates when $\bar{F}_{\rm in}$ is of order $10^{23}$ years \cite{astashenok12}.

As another example, consider the PR model (\ref{17}). The inertial force (\ref{24}) is equal to \cite{frampton12}
\begin{equation}
F_{\rm in}=ml\left[ \lambda H_1e^{-\lambda t}+(H_0-H_1e^{-\lambda t})^2\right]. \label{35}
\end{equation}
In this case the inertial force is limited: $F_{\rm in}\rightarrow mlH_0^2$ when $t \rightarrow +\infty$.

As the time $t$ increases, the universe develops to an expanding de Sitter space-time determined by the parameter $\Lambda_0$. Choose the relationship between $w_0$ and $\Lambda_0$ in equation (\ref{5}) in the form
\begin{equation}
w_0=-1-\frac{\kappa^2\Lambda_0}{3H_0^2}. \label{36}
\end{equation}
Then the energy conservation law is written as
\begin{equation}
\kappa^2\Lambda_0\frac{H_1}{H_0}e^{-\lambda t}+2\left(\lambda-\frac{\kappa^2\Lambda_0}{H_0}\right)=0, \label{37}
\end{equation}
and one finds for the Rip time $t_{PR}$
\begin{equation}
t_{PR}=-\frac{1}{\lambda}\ln \frac{2H_0}{H_1}\left( 1-\frac{\lambda H_0}{\kappa^2\Lambda_0}\right). \label{38}
\end{equation}
It is thus possible, from the relations (\ref{34}) and (\ref{38}), to estimate the span of time $t_{LR}$ or $t_{PR}$ needed before the system becomes gravitationally unbound.

\section{Conclusions}

In summary, we have presented dark energy models with an inhomogeneous equation of state, cf. equation (\ref{5}) above, in which Little Rip or Pseudo-Rip behavior is encountered in the far future.  We have shown that the LR cosmology is caused exponentially, determined by the parameters $\Lambda$ or  $w$. It is of interest to note that the disintegration of bound structures in the LR/PR models may occur for physically acceptable choices of parameters. We have given estimates for the disintegration times, in case of the standard model for LR, and also for the case of the asymptotic de Sitter universe.

As a final comment, we mention that it is possible to generalize the LR  theory so as to take into account the  bulk viscosity in the dark fluid. Such a formulation was developed in \cite{brevik11}. One convenient choice for the equation of state, which permits an analytic solution to be found, is the following,
\begin{equation}
p=-\rho-A\sqrt{\rho}-\xi_0. \label{39}
\end{equation}
Here $A$ is a positive constant (with dimension cm$^{-2}$ in geometric units), and the last term $\xi_0$ is a constant representing the influence from the bulk viscosity. Some calculation (details omitted here) shows that $\rho(t)$ increases exponentially for large values of $t$. This is the characteristic feature for the LR again, now with the inclusion of viscosity.

\vspace{1.5cm}

{\bf Acknowledgement}.  We thank Professor Sergei Odintsov for very useful discussions.

\end{document}